\documentclass[aps, twocolumn, prl, preprintnumbers,superscriptaddress]{revtex4-1}

 \newcommand{\mytitle}[1]{
 \twocolumn[\hsize\textwidth\columnwidth\hsize
 \csname@twocolumnfalse\endcsname #1 \vspace{1mm}]}
 \newcommand{\beq}{\begin{equation}}
 \newcommand{\eeq}{\end{equation}}
 \newcommand{\bea}{\begin{eqnarray}}
 \newcommand{\eea}{\end{eqnarray}}

\usepackage{amsmath,amsfonts,amssymb,graphics,graphicx,epsfig,color,times,indentfirst,
layout,hyperref,multirow,bm}
\usepackage{hyperref}
\usepackage{mathtools}
\usepackage{bbold}
\usepackage{ulem}
\hypersetup{linktocpage,colorlinks=true,citecolor=blue}

\usepackage{soul}

\usepackage{graphicx}
\usepackage{dcolumn}

\usepackage{xcolor}
\usepackage{mathrsfs}
\usepackage{fixmath}
\usepackage{subfloat}

\usepackage[all]{hypcap} 

\newcommand{\mbd}{\mathbold}

 \usepackage[caption=false]
 {subfig}
 \usepackage{lipsum}
 \usepackage{titlesec}
  \usepackage{fancyhdr}
  
\begin{document}

\title{Universal Non-equilibrium $\mbd{I}$-$\mbd{V}$ Curve Near Two-channel Kondo-Luttinger Quantum Critical Point}

\author{C.-Y. Lin}
\affiliation{Department of Electrophysics, National Chiao-Tung University, Hsinchu, 30010
Taiwan, R.O.C.}
\affiliation{Institute of Physics, Academia Sinica, Taipei, 11529, Taiwan, R.O.C.}
\author{Y.-Y. Chang}
\affiliation{Department of Electrophysics, National Chiao-Tung University, Hsinchu, 30010 Taiwan, R.O.C.}
\author{C. Rylands}
\affiliation{Department of Physics, Rutgers University, Piscataway, New Jersey, 08854, U.S.A.}
\affiliation{Joint Quantum Institute, University of Maryland, College Park, Maryland, 20742, U.S.A.}
\affiliation{Condensed Matter Theory Center, University of Maryland, College Park, Maryland, 20742, U.S.A.}
\author{N. Andrei}
\affiliation{Department of Physics, Rutgers University, Piscataway, New Jersey, 08854, U.S.A.}
\author{C.-H. Chung}
\affiliation{Department of Electrophysics, National Chiao-Tung University, Hsinchu, 30010 Taiwan, R.O.C.}
\affiliation{Institute of Physics, Academia Sinica, Taipei, 11529, Taiwan, R.O.C.}

\date{\today}

\begin{abstract}
The Fermi liquid paradigm for metals has contributed enormously to our understanding of condensed matter systems. However a growing number of quantum critical systems have been shown to exhibit non-Fermi liquid behavior. A full understanding of such systems is still lacking and in particular analytical results away from  equilibrium are rare. In this work, we provide a distinct example of such kind in a two-channel Kondo-Luttinger model where a Kondo impurity couples to two voltage-biased interacting electron leads, experimentally realizable in a dissipative quantum dot. Since the 1990's, an exotic quantum phase transition has been known to exist from the 1-channel to 2-channel Kondo ground states by enhancing electron interactions in the leads, but a controlled theoretical approach to this quantum critical point has not yet been established. We present a controlled method to this problem and obtain an analytical form for the universal non-equilibrium differential conductance near the transition. The relevance of our results for recent experiments is discussed.
\end{abstract}
\maketitle

\textit{\textbf{Introduction.}}---Over the recent decades, there has been growing experimental evidence for correlated electron systems whose low temperature thermodynamic and transport properties violate Landau's Fermi liquid paradigm for metals \cite{HewsonBook, FLBook}. Such non-Fermi liquid (NFL) behavior, ranging from heavy-fermion, unconventional superconductors \cite{NFL-HF, 115, Taillefer} to Kondo impurity quantum dot systems \cite{KondoPC, 2ck-Goldhaber}, often appears near a quantum phase transition (QPT) \cite{SachdevBook} as a result of competing ground states. Despite their prevalence, very few examples exist where analytical or exact NFL results are available \cite{Sela, Duke-NCTU}. While the equilibrium aspect of QPTs has been extensively studied, much less is known however about out of equilibrium quantum critical properties, more relevant for experiments, either by a voltage bias \cite{ChungPRL09, ChungPRB13, Sela, Duke-NCTU} or by a sudden quantum quench \cite{quench-QPT}. Highly tunable nano-scale quantum impurity systems offer an excellent playground to study NFL near non-equilibrium QPTs \cite{Goldhaber-Gordon QD, JJLin}.
  
  One example of such systems is the Kondo-Luttinger model, experimentally realizable in a dissipative Kondo dot device \cite{Mebrahtu12, Mebrahtu13}. Therein a spin-$1/2$ Kondo impurity at the center couples to two (left $L$ and right $R$) Luttinger liquid wires of a total length $\mathcal{L}$ with repulsive electron-electron interactions, giving a Luttinger parameter $K<1$ (see Fig. 1(a)). This model introduces inter-lead (intra-lead) Kondo couplings $J_{LR}$ ($J_{LL/RR}$), involving screening of the impurity spin by the conduction electrons of the both leads (one lead), respectively. In the weak-coupling limit ($J_{\alpha\alpha'}\rightarrow0$) at a higher temperature, electron repulsive interactions in the leads are known to suppress the $J_{LR}$ terms in a $T$-power-law fashion: $J_{LR}\sim T^{1/2(1/K-1)}$ \cite{GogolinPRB1995, E.Kim}; while $J_{LL/RR}$ terms are unaffected by interactions and show a $T$-logarithmic divergence of a typical Kondo effect. Since mid-1990's it has been predicted that with increasing electron interaction (or decreasing $K$) and $T\rightarrow0$, this model supports an exotic QPT from the conducting 1-channel Kondo (1CK) ground state to the insulating 2-channel Kondo (2CK) ground state. Wherein the 1CK ground state, both $J_{LR}$ and $J_{LL/RR}$ couplings are $T$-power law divergent and the two leads are coupled to form a single Kondo screening channel; in the 2CK ground state, the $J_{LR}$ ($J_{LL/RR}$) coupling is $T$-power law suppressed (enhanced) and the two leads independently Kondo-screen the impurity spin \cite{Furusaki94, GogolinPRB1995}. The 1CK-2CK quantum critical point (QCP) is expected at $K=1/2$ \cite{Furusaki94, GogolinPRB1995}. However, accessing the NFL properties of this QCP becomes challenging due to the lack of controlled theoretical approaches to physics near the strong coupling 2CK ground state where $J_{LL/RR}\rightarrow\infty$ and the standard perturbation theory fails. 

  In this letter, we re-examine the Kondo-Lutinger system and establish a controlled theoretical framework to circumvent the above difficulty. We address the non-equilibrium transport near $K=1/2$ via a bosonization-refermionization approach \cite{ChungNJP2015} combined with the Keldysh Green's function method. This approach maps the strong coupling problem onto an effective weak coupling one where controlled many-body technique is applicable. The Hamiltonian is first bosonized \cite{Giamarchibook, Gogolinbook}, followed by the Emery-Kivelson transformation \cite{E-K Trans}, which moves the system to near the strong coupling 2CK fixed point. The Hamiltonian near 2CK is then expressed in terms of the leading irrelevant inter- and intra-lead Kondo couplings in the weak coupling regime. To carry out the calculations on charge transport, the Hamiltonian near 2CK fixed point is further re-fermionized as two effective voltage-biased free fermion leads coupled to an impurity spin and a bosonic bath. Since the current is determined by $J_{LR}$, we study the renormalization group (RG) flow of this coupling and find the QCP, which allows for a reliable study in the weak coupling regime. To simplify our calculations, we work in the channel symmetric case ($J_{LL}=J_{RR}$) and near the Toulouse limit where only $J_{LR}$ dominates. Nevertheless, our results can be extended more generally to parameter space away from Toulouse limit. This can be done for the following reasons: (i) the operators around this limit--the transverse ($xy$) component of $J_{LR}$ and the longitudinal ($z$) component of $J_{LL/RR}$--are all irrelevant and hence will always stay in the weak coupling regime, and (ii) the RG flow for $J_{LR}$ at 1-loop order in this limit via Eq. (\ref{eq:RG-eq}) shows a negligible difference from that up to 2-loop order and away from this limit (see Inset of Fig. 1(c)). This shows that our analytic results based on Eq. (\ref{eq:RG-eq}) is accurate and reliable enough to be extended to the parameter space away from Toulouse limit. The universal non-linear $I$-$V$ curve of the effective model is analytically obtained near QCP for $K<1/2$ via the Keldysh non-equilibrium Green's function formalism. The interactions in the leads are treated exactly by bosonization, and the current is computed perturbatively in $J_{LR}$.  The main point of this work is not the RG analysis of the Hamiltonian but the actual calculations of non-equilibrium current near criticality. Our results offer an unique example of analytically accessible non-equilibrium transport near an impurity quantum critical point.  

\textit{\textbf{The Kondo-Luttinger model.}}---The Hamiltonian of our system in the presence of particle-hole symmetry reads \cite{Giamarchibook, Gogolinbook}  $H=H_0+H_{int}+H_K+H_{\mu}$: $H_0=-iv_F\sum_{\alpha;\sigma}\int dx\left[\mathfrak{R}^{\dagger}_{\alpha,\sigma}(x)\partial_x \mathfrak{R}_{\alpha,\sigma}(x)-\mathfrak{L}^{\dagger}_{\alpha,\sigma}(x)\partial_x \mathfrak{L}_{\alpha,\sigma}(x)\right]$, $H_{int}=\sum_{\alpha;\sigma,\sigma'}
\int dx\left[\frac{g_{4}}{2}(\rho_{\alpha,\sigma}(x)\rho_{\alpha,\sigma'}(x)+\overline{\rho}_{\alpha,\sigma}(x)\overline{\rho}_{\alpha,\sigma'}(x)\right.$\\$\left.)+g_{2}\rho_{\alpha,\sigma}(x)\overline{\rho}_{\alpha,\sigma'}(x)\right]$, $H_K=\sum_{i;\alpha,\alpha';\sigma,\sigma'} J_{\alpha\alpha'}S_i\cdot\psi^{\dagger}_{\alpha,\sigma}(0)$\\$\times\frac{\tau^{i}_{\sigma,\sigma'}}{2}\psi_{\alpha',\sigma'}(0)$, $H_{\mu}=\frac{eV}{2}\sum_{\alpha,\sigma}\int dx\left[\rho_{\alpha,\sigma}(x)+\overline{\rho}_{\alpha,\sigma}(x)\right]$, where $\alpha=L,R$, $\sigma=\uparrow,\downarrow$ are the lead and spin indices respectively, $S_i$ is the impurity spin and $\tau^i_{\sigma,\sigma'}$ is the Pauli matrix with $i=x$, $y$ or $z$, and we set $\hbar=1$. The integrations are taken from $x=-\mathcal{L}/2$ to $x=0$ for $\alpha=L$, and from $x=0$ to $x=\mathcal{L}/2 $ for $\alpha=R$. The electron field operator is defined as: $\psi_{\alpha,\sigma}(x)=\mathfrak{R}_{\alpha,\sigma}(x)+\mathfrak{L}_{\alpha,\sigma}(x)$, with $\mathfrak{R}_{\alpha,\sigma}(x)$ ($\mathfrak{L}_{\alpha,\sigma}(x)$) being the right (left) moving electrons; the corresponding electron density operators are $\rho_{\alpha,\sigma}(x)=\mathfrak{R}^{\dagger}_{\alpha,\sigma}(x)\mathfrak{R}_{\alpha,\sigma}(x)$ and $\overline{\rho}_{\alpha,\sigma}(x)=\mathfrak{L}^{\dagger}_{\alpha,\sigma}(x)\mathfrak{L}_{\alpha,\sigma}(x)$. Here, $H_0+H_{int}$ describes the Luttinger liquid wire with $H_0$ being free electron leads and $H_{int}$ the electron-electron interactions in the leads, $H_{Kf}$ and $H_{Kb}$ describe the Kondo couplings, $H_{\mu}$ is the bias voltage term.

   Near the weak-coupling fixed point ($J_{\alpha\alpha'}\rightarrow0$) at a higher energy scale $T\sim D$ (with $D$ being the bandwidth), it is convenient to represent $H$ in terms of chiral boson fields through standard bosonization: $\Psi_{\alpha,\sigma}(x)=\lim_{a\rightarrow0}\frac{1}{\sqrt{2\pi a}}\eta_{\alpha,\sigma}e^{-i\phi_{\alpha,\sigma}(x)}$ with $\eta_{\alpha,\sigma}$ the Klein factor and $\phi_{\alpha,\sigma}(x)$ the chiral boson fields \cite{Gogolinbook, suppl}. To explore the physics near the strong-coupling regime for $J_{\alpha\alpha}\rightarrow\infty$ as the system approaches to the ground, the Emery-Kivelson transformation \cite{E-K Trans} is then performed on the chiral bosonized Hamiltonian, $H_{cb}$: $U^{\dagger}H_{cb} U=H_{sc}+H_{\mu}$ with $U=e^{-iS_z\phi_s(0)}$ \cite{GogolinPRB1995}, where   
  \begin{align}
  H_{sc}=&\int^{\frac{\mathcal{L}}{2}}_{-\frac{\mathcal{L}}{2}}\frac{dx}{4\pi}
  \Big(\sum_{\mathclap{\mu=c,f}}v_c[\nabla\phi_{\mu}(x)]^2
  +\sum_{\mathclap{\nu=s,sf}}\;v_F[\nabla\phi_{\mu}(x)]^2\Big)\nonumber
  \\
  +&\dfrac{J_{LR}}{\pi a}S_x\cos\left(\frac{\phi_{f}}{\sqrt{K}}\right)
  -\frac{J_{LR}^z}{\pi a}S_z\sin\phi_{sf}\sin\left(\frac{\phi_{f}}{\sqrt{K}}\right)\nonumber
  \\
  +&\dfrac{J_+}{\pi a}S_x\cos\phi_{sf}-\dfrac{J_-}{\pi a}S_y\sin\phi_{sf}\nonumber
  \\
  +&\dfrac{\left(J^z_+-2\pi v_F\right)}{4\pi}S_z\nabla\phi_s
  +\frac{J_-^z}{4\pi}S_z\nabla\phi_{sf},
  \nonumber\\
  H_{\mu}=&\frac{eV}{4\pi}\int^{\frac{\mathcal{L}}{2}}_{-\frac{\mathcal{L}}{2}}dx
  \nabla\phi_{f}(x),
  \label{eq:KL-hamiltonian}
\end{align}
where the chiral boson fields $\phi_{c/f/s/sf}$ are defined as  $\phi_c=\sum_{\alpha=L}^R\phi_{c\alpha}/\sqrt{2}$, $\phi_f=\sum_{\alpha=L}^R\tau^z_{\alpha,\alpha}\phi_{c\alpha}/\sqrt{2}$, $\phi_s=\sum_{\alpha=L}^R\phi_{s\alpha}/\sqrt{2}$, $\phi_{sf}=\sum_{\alpha=L}^R\tau^z_{\alpha,\alpha}\phi_{s\alpha}/\sqrt{2}$, and $\phi_{c\alpha}=\sum_{\sigma}\phi_{\alpha,\sigma}/\sqrt{2}$, $\phi_{s\alpha}=\sum_{\sigma=\uparrow}^{\downarrow}\tau^z_{\sigma,\sigma}\phi_{\alpha,\sigma}/\sqrt{2}$ \cite{E-K Trans}. $J_+=\frac{J^{\perp}_{LL}+J^{\perp}_{RR}}{2}$, $J_-=\frac{J^{\perp}_{LL}-J^{\perp}_{RR}}{2}$, $J^z_+=\frac{J^z_{LL}+J^z_{RR}}{2}$, $J^z_-=\frac{J^z_{LL}-J^z_{RR}}{2}$. In addition, $v_c$ is the renormalized Fermi velocity and $K \equiv \sqrt{\frac{1- g_2/(8\pi v_F + g_4 )}{1+ g_2/(8\pi v_F + g_4)}}$ \cite{Gogolinbook, suppl}.

  Note that $J_{+}$, $J_{-}$ become the most relevant (with a scaling dimension $[J_{+/-}]=1/2$), while $J_{LR}$ is the leading irrelevant term for $K<1/2$ with $[J_{LR}]=1/2K$ and hence remains in the weak-coupling regime. The more irrelevant terms are $J^z_{+/-}$ ($[J^z_{+/-}]=1/2K+1$) and $J^z_{LR}$ ($[J^z_{LR}]=1/K+1/2$). In the following, we discuss the model (Eq. (\ref{eq:KL-hamiltonian})) in the channel symmetric case ($J^{\perp}_{LL}=J^{\perp}_{RR}$ and $J^z_{LL}=J^z_{RR}$) so the channel asymmetric terms $J_-$, $J^z_{-}$ are absent, and near the Toulouse limit ($\delta J^z=J^z_{+}-2\pi v_{F}\ll\mathcal{O}(1)$). Also, the most relevant $J_+$ term is pinned at a large value, while the most irrelevant $J^z_{LR}$ term is neglected here. As a result, only the leading irrelevant Kondo couplings $J_{LR}$ and $\delta J^z$ terms survive. 

  For $K\leq1/2$, it is useful for later analysis to re-fermionize Eq. (\ref{eq:KL-hamiltonian}) near the strong coupling 2CK fixed point and the Toulouse limit in terms of effective free fermions weakly coupled to an impurity spin and an Ohmic bosonic bath $H'_{b}=\frac{v_c}{4\pi}\int(\nabla\varphi(x))^2dx$. With the following transformation \cite{ChungNJP2015}:
  \begin{align}
  \left\{
  \begin{array}{l}
  \sqrt{\dfrac{1}{K}}\phi_f=\sqrt{2}\phi'_f+\sqrt{\dfrac{1}{K}-2}\varphi
  \\
  \\
  \sqrt{\dfrac{1}{K}}\widetilde{\phi}=\sqrt{\dfrac{1}{K}-2}\phi'_f-\sqrt{2}\varphi
  \end{array}
  \right.,
  \label{eq:transformation}
  \end{align}
the Hamiltonian Eq. (\ref{eq:KL-hamiltonian}) can be further re-fermionized as $H_{sc}^\prime+H^\prime_{\mu}+H^\prime_b=H_{sc}+H_{\mu}+H_b $, where  
  \begin{align}
  H_{sc}'&=\sum_{\mathclap{\substack{\mu=cL,cR;\\k}}}
  v_ck\;c{'}^{\dagger}_{\mu,k}c{'}^{}_{\mu,k}
  +\sum_{\mathclap{\substack{\nu=sL,sR,\\k}}}
  v_Fk\;c{'}^{\dagger}_{\nu,k}c{'}^{}_{\nu,k}
  +\dfrac{J_+}{\pi a}S_x\cos\phi_{sf}
  \nonumber\\
  &+\dfrac{J_{LR}}{\mathcal{L}}S_x\sum_{k,k'}\left(c{'}^{\dagger}_{cL,k}c{'}^{}_{cR,k'}
  e^{i\sqrt{\frac{1}{K}-2}\varphi(0)}+H.C.\right)
  \nonumber\\
  &+\dfrac{\delta J^z}{\sqrt{2}\mathcal{L}}S_z\sum_{k}
  \left(c{'}^{\dagger}_{sL,k}c^{}{'}_{sL,k}+c{'}^{\dagger}_{sR,k}c^{}{'}_{sR,k}\right),
  \nonumber\\
  H_{\mu}'&=eV\sqrt{K}\sum_{k}
  \left(c{'}^{\dagger}_{cL,k}c^{}{'}_{cL,k}+c{'}^{\dagger}_{cR,k}c{'}^{}_{cR,k}\right)
  \nonumber\\
  &+eV\sqrt{1-2K}\int^{\frac{\mathcal{L}}{2}}_{-\frac{\mathcal{L}}{2}}\frac{dx}{4\pi}\nabla\varphi(x),
  \nonumber\\
  H'_b&=\dfrac{v_c}{4\pi}\int^{\frac{\mathcal{L}}{2}}_{-\frac{\mathcal{L}}{2}}(\nabla\varphi(x))^2dx.
  \label{eq:refermionized-H}
  \end{align}
Here, the $k$-space effective free fermions in this new basis reads: $c^\prime_{\mu,k}=\frac{1}{\sqrt{\mathcal{L}}}
\int^{\frac{\mathcal{L}}{2}}_{-\frac{\mathcal{L}}{2}}\Psi^\prime_{\mu}(x)
e^{-ikx}dx$ with $\Psi^\prime_{cL/cR}(x)=\lim_{a \rightarrow 0}\frac{1}{\sqrt{2\pi a}}\eta_{cL/cR}e^{-i(\phi_c(x)\pm \phi^\prime_f(x))/\sqrt{2}}$, and $\Psi^\prime_{sL/sR}(x)=\Psi_{sL/sR}(x)=\lim_{a\rightarrow0}\frac{1}{\sqrt{2\pi a}}\eta_{sL/sR}e^{-i(\phi_s(x)\pm\phi_{sf}(x))/\sqrt{2}}$. Eq. (\ref{eq:refermionized-H}) is an effective weak coupling Hamiltonian ($J_{LR}$, $\delta J^z<1$) near strong coupling ($J_+ \rightarrow \infty$) 2CK fixed point where standard perturbation theory is applicable. It describes two voltage-biased free fermion leads ($c'_{cL/cR}$) showing an inter-lead coupling to an impurity spin ($S_x$) subject to a dissipative baonic bath; while another two free fermion leads ($c'_{sL/sR}$) couple to $S_z$. 

\begin{figure}[htp]

  \subfloat{\includegraphics[clip,width=0.5\columnwidth]{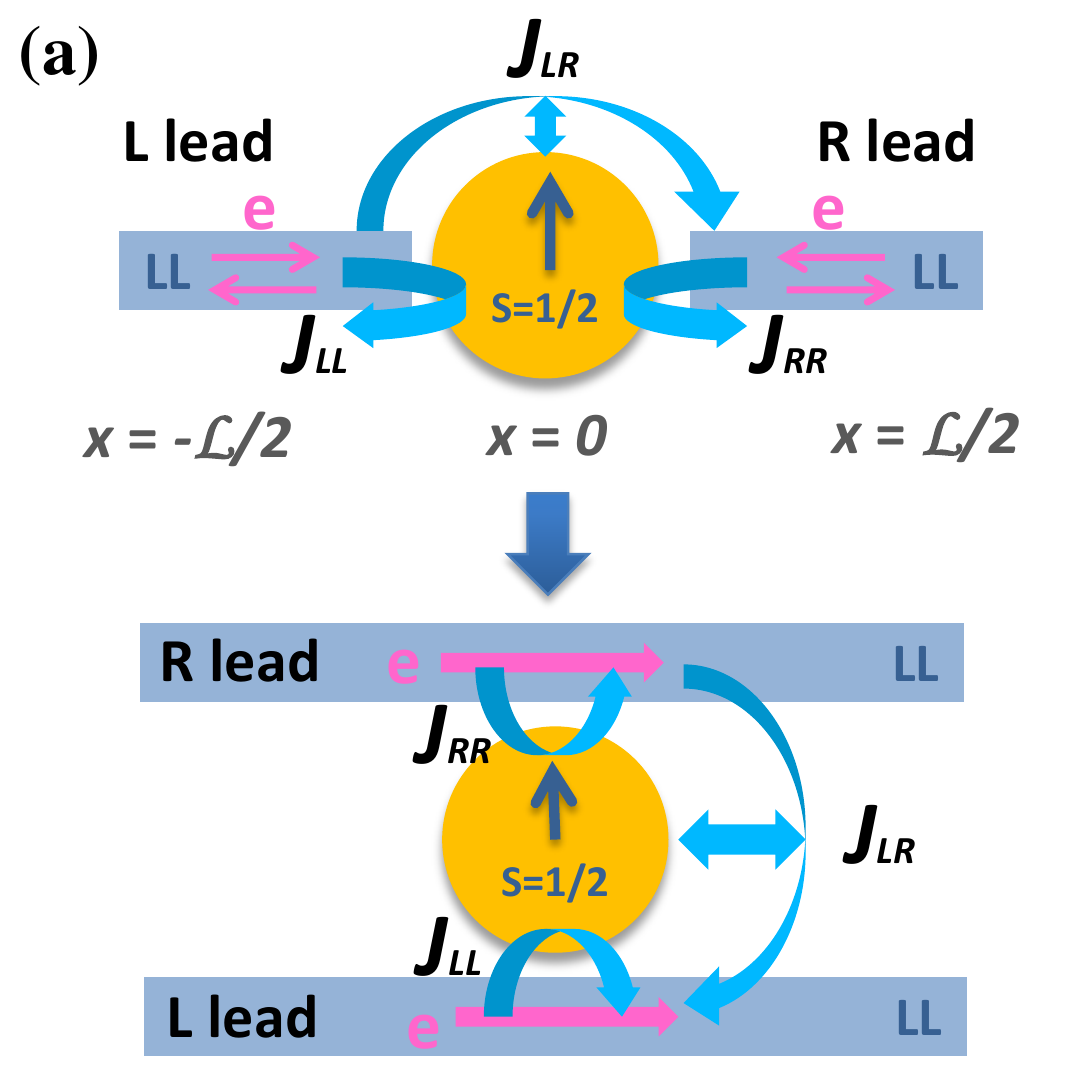}}
  \subfloat{\includegraphics[clip,width=0.5\columnwidth]{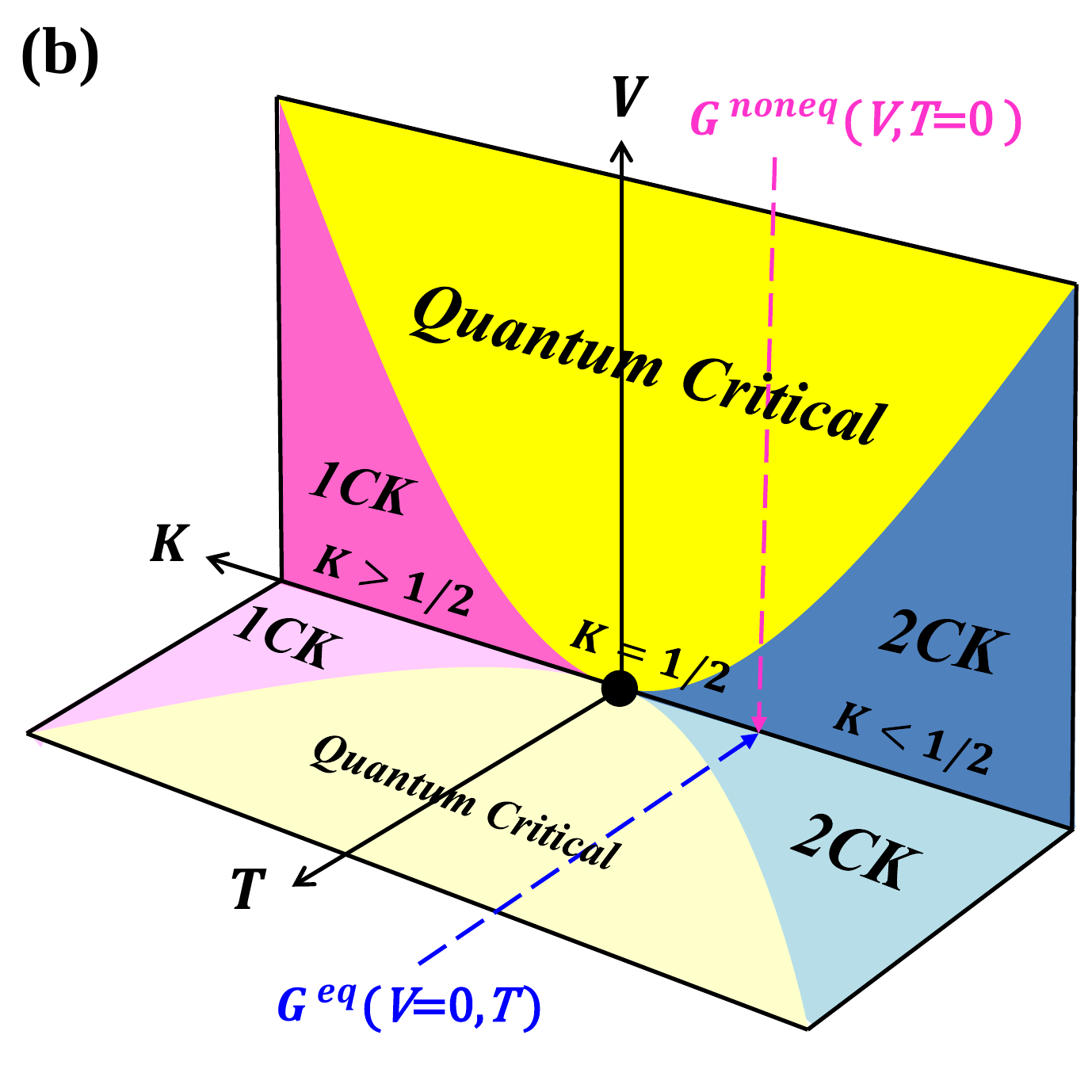}}
  
  \subfloat{\includegraphics[clip,width=0.5\columnwidth]{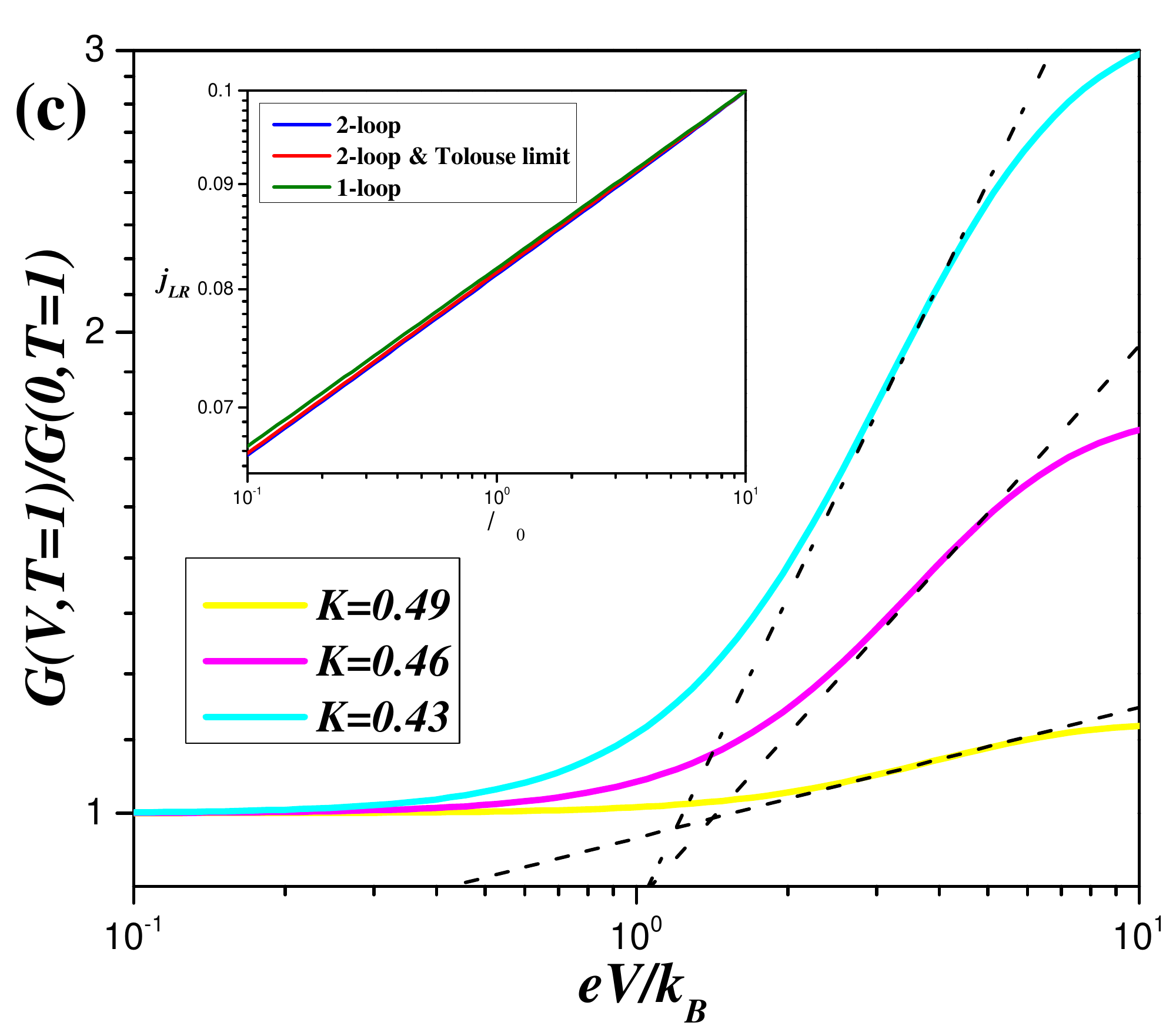}}
  \subfloat{\includegraphics[clip,width=0.5\columnwidth]{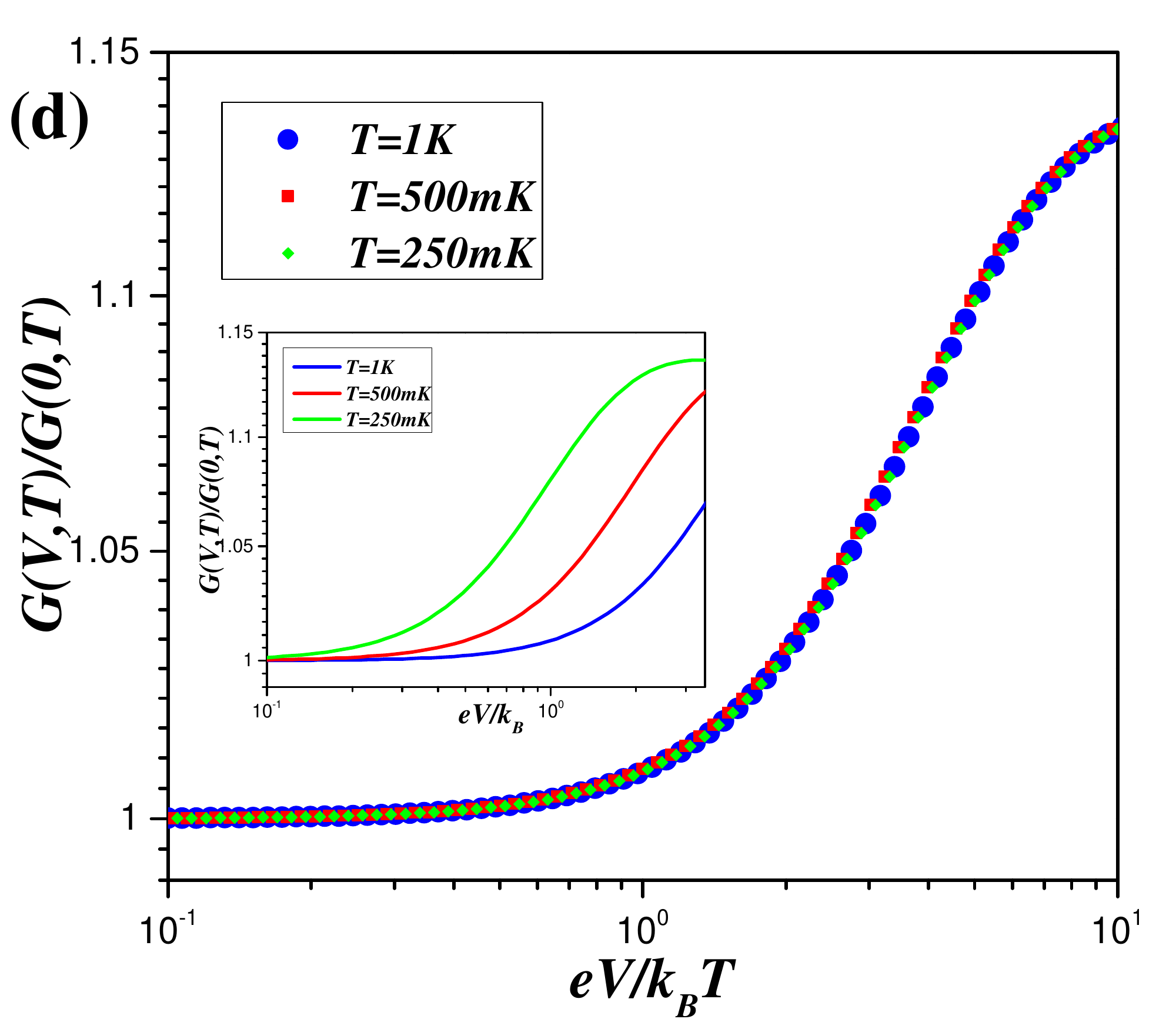}}

\caption{\small (a) The original Kondo-Luttinger model (above) with two electron branches (left moving and right moving) in each of the two leads of a length $\mathcal{L}/2$ can be transformed to an equivalent chiral Kondo-Luttinger model (below) where both leads are now unfolded to extend from $-\mathcal{L}/2$ to $\mathcal{L}/2$ with only one electron branch left. (b) Schematic phase diagram of the Kondo-Luttinger model as functions of $V$, $T$ and the Luttinger parameter $K$. (c) $G(V,\,T=1\text{K})/G(0,\,T=1\text{K})$, the normalized non-equilibrium differential conductance at the strong coupling fixed point with different values of Luttinger parameter $K\leq1/2$. The dashed lines are power-law fit to $V^{1/K-2}$. Inset: RG flows for $j_{LR}$ for $K=0.46$ with bare coupling $j_{LR}=0.1$ up to 1-loop order (green), 2-loop in Toulouse limit (red), 2-loop away from Toulouse limit with bare value of $\delta j^z=0.2$ (blue). (d) Universal $eV/k_BT$ scaling in normalized differential conductance $G(V,\,T)/G(0,\,T)$ at various temperatures for $K=0.49$. The Inset shows non-rescaled conductances.}
\label{fig:fig1}
  \end{figure}

  Since the charge transport is determined by $J_{LR}$, we further perform RG analysis. The RG scaling equations up to one-loop order for $J_{LR}$ and $\delta J^z$ are derived via Eq. (\ref{eq:refermionized-H}) as \cite{suppl}
  \begin{align}
  \dfrac{dJ_{LR}}{dl}=\left(1-\dfrac{1}{2K}\right)J_{LR},
  ~~\dfrac{d\delta J^z}{dl}=-\dfrac{1}{2K}\delta J^z,
  \label{eq:RG-eq}
  \end{align}
where $dl=-\frac{d\Lambda}{\Lambda}$ with $\Lambda$ being a running energy cut-off. Note that we find no contributions quadratic in Kondo couplings to Eq. (\ref{eq:RG-eq}) due to decoupling of the fields $c_{cL/cR}$ in $J_{LR}$ term from $c_{sL/sR}$ in $\delta J^z$ term. Based on Eq. (\ref{eq:RG-eq}), $J_{LR}$ term is irrelevant (relevant) for $K<1/2$ ($K>1/2$). As a result, the 1CK-2CK QCP occurs at $K=1/2$, separating 1CK state with $J_+\rightarrow\infty$, $J_{LR}\rightarrow\infty$ for $K>1/2$ from the 2CK state with $J+\rightarrow\infty$, $J_{LR}\rightarrow0$ for $K<1/2$ (see Fig. \ref{fig:fig1}(b)) \cite{GogolinPRB1995}.
  
\textit{\textbf{Non-equilibrium charge current near QCP.}}---The equilibrium transport of our model is known \cite{GogolinPRB1995, E.Kim}: with decreasing temperatures from the weak-coupling ($J_{\alpha\alpha'}\rightarrow0$) fixed point at $T=D$, the Hamiltonian $H$ gives a $T$-power-law suppressed differential conductance $G(T)\sim T^{1/K-1}$ via $[J_{LR}]=(1+1/K)/2$; while as $T\rightarrow0$, it shows a different $T$-power-law near QCP in the strong coupling limit ($J_+\rightarrow\infty$) via Eq. (\ref{eq:RG-eq}): $G(T)\sim T^{1/K-2}$ (see Fig. \ref{fig:fig1}(b) and the blue dashed arrow therein). However, little is known about the non-equilibrium transport at a finite voltage bias near QCP (see Fig. \ref{fig:fig1}(b) and the red dashed arrow therein), which is expected to show features distinct from its equilibrium counterpart.
  
  Near QCP, the steady state charge current, defined as the charges passing through the Kondo dot from the left to right lead per unit time, is derived from the Heisenberg equation of motion via Eq. (\ref{eq:refermionized-H}): 
  \begin{align}
  I&=-e\frac{d\langle N_L\rangle}{dt}=\frac{\sqrt{2}ie}{\hbar}
  \langle[\hat{N}_{cL},H_{sc}]\rangle
  \nonumber\\
  =&\sqrt{2K}\left(3-\dfrac{1}{2K}\right)\frac{e}{\hbar}\frac{J_{LR}}{\mathcal{L}}\langle 
  S_x\rangle\sum_{k,k'}\text{Re}\{G^<_{RL,k'k}(t,t)\},
  \label{eq:current}
  \end{align}
where $N_L=N_{L\uparrow}+N_{L\downarrow}=\sqrt{2}N_{cL}$ is obtained via  $N_{cL}=(1/4\pi)\int\nabla\phi_{cL}(x)dx$, and $G^<_{RL,k'k}(t,t')=i\langle c{'}^{\dagger}_{cL,k}(t')c{'}^{}_{cR,k'}(t)e^{i\sqrt{\frac{1}{K}-2}\varphi(t)}\rangle$ is the lesser Green's function for $c_{cL/cR,k}$.

  Within the Keldysh non-equilibrium Green's function approach \cite{Haugbook, MeirWingreenPRL92}, we first derive the equation of motion for the contour-ordered Green's function $G^c_{RL,k'k}(\tau-\tau')$, expressed as the Dyson type integral equation \cite{suppl}: 
  \begin{align}
  G^c_{RL,k'k}(\tau-\tau')=\dfrac{J_{LR}}{\mathcal{L}}S_x\sum_{k''}\int &dt_1
  \left[G^c_{R,k'k''}(\tau-\tau_1)\right. 
  \nonumber\\  
  &\left.\times g^c_{L,k}(\tau_1-\tau')\right],
  \end{align} 
where $G^c_{R,k'k''}(\tau-\tau')=-i\langle T_c[c{'}^{}_{cR,k'}(\tau)e^{i\sqrt{\frac{1}{K}-2}\varphi(\tau)}$ $\times c{'}^{\dagger}_{cR,k''}(\tau')e^{-i\sqrt{\frac{1}{K}-2}\varphi(\tau')}]\rangle$ and $g^c_{L,k}(\tau-\tau')=-i\langle T_t[c{'}_{cL,k}(\tau) c{'}^{\dagger}_{cL,k}(\tau')]\rangle_0$ is the free fermion contour-ordered Green's function. 
The time-lesser Green's function $G^<_{RL,kk'}(t-t')$ expanded perturbatively up to first-order in $J_{LR}$, is given by \cite{suppl}: 
  \begin{align}
  &G^<_{RL,k'k}(t-t')
  \nonumber\\
  =&\dfrac{J_{LR}}{\mathcal{L}}S_x\sum_{k''}\int^{\infty}_{-\infty} dt_1
  [g^r_{R,k'}(t-t_1)g^<_{L,k}(t_1-t')
  \nonumber\\
  &~~~~~~~~~~~~~~~~~~~+g^<_{R,k'}(t-t_1)g^a_{L,k}(t_1-t')]b^<(t-t_1)
  \nonumber\\
  +&[g^r_{R,k'}(t-t_1)+g^<_{R,k'}(t-t_1)]g^<_{L,k}(t_1-t')b^r(t-t_1),
  \label{eq:1st ord green}
  \end{align}
where $g^{r/a}_{R/L,k}$, $g^<_{R/L,k}$ are retarded/advanced and lesser component of bare Green's functions of the effective non-interacting right/left lead, respectively, and 
$b^<(t-t')=-i\langle e^{-i\sqrt{\frac{1}{K}-2}\varphi(t')}e^{i\sqrt{\frac{1}{K}-2}\varphi(t)}
\rangle_0$, $b^r(t-t')=-i\theta(t-t')\times\langle[e^{-i\sqrt{\frac{1}{K}-2}\varphi(t')}, e^{i\sqrt{\frac{1}{K}-2}\varphi(t)}]\rangle_0$ are the lesser and retarded bosonic correlation functions calculated with respect to $H'_b$ only. 

  In the thermodynamic limit, the explicit analytical form of the non-equilibrium current via Eqs. (\ref{eq:current}) and (\ref{eq:1st ord green}) reads \cite{suppl}
  \begin{align}
  I=&\dfrac{\sqrt{2K}}{4}\left(3-\dfrac{1}{2K}\right)\frac{e}{\hbar}
  j^2_{LR}\langle S_x\rangle\int^{\infty}_{-\infty}~~
  \mathclap{d\varepsilon_{k}d\varepsilon_{k'}}~~~~~~[f(\varepsilon_{k'}-eV/2)
  \nonumber\\
  &-f(\varepsilon_{k}+eV/2)]
  \times\int^{\infty}_{-\infty}dt b(t)e^{\frac{i}{\hbar}(\varepsilon_{k'}-\varepsilon_{k})t}
  \nonumber\\
  =&-\dfrac{\sqrt{2K}}{4}\left(3-\dfrac{1}{2K}\right)\frac{e}{\hbar}j_{LR}^2
  \langle S_x\rangle
  \dfrac{V}{\Gamma(\frac{1}{K}-1)}\left(\frac{2\pi k_BT}{\hbar D}\right)^{\frac{1}{K}-2}
  \nonumber\\
  &\times\left|\frac{\Gamma(\frac{1}{2K}+i\frac{eV/2}{2\pi k_BT})}{\Gamma(1+i\frac{eV/2}
  {2\pi k_BT})}\right|^2,
  \label{I}
  \end{align}
where $G^<_{RL,kk'}(\omega)$ is the right-left lesser Green's function in the frequency space, $j_{LR}=J_{LR}/2\pi\hbar v_F$, $2D$ is the bandwidth of the bosonic bath $H_b'$, and $\Gamma(x)$ is the Gamma function. Note that the non-linear transport of our system in the Toulouse limit for $K=1$ (with non-interacting leads) was addressed in Ref. \cite{SchillerHershToulousePRB98}. There, the $J_{LR}$ term is relevant and the system goes to a resonant-tunneling ground state with the quantum unitary conductance $G=2e^2/h$ \cite{SchillerHershToulousePRB98}.

  
  The analytical differential conductance $G(V,T)=dI/dV$ near QCP for $K<1/2$ via Eq. (\ref{I}) is plotted in Figs. \ref{fig:fig1}(c) and \ref{fig:fig1}(d) for various values of $K$ and temperatures. Near QCP, the equilibrium conductance shows a $T$-power-law suppression: $G(V=0,T)\propto T^{1/K-2}$ \cite{FlorensPRB2007}, leading to an insulating 2CK state. However, for $V>T$ and for a fixed $T=T_0$, $G(V,T_0)$ deviates from the power-law $V^{1/K-2}$ (dashed lines in Fig. \ref{fig:fig1}(c)), and the deviation becomes larger at large bias $V\gg T_0$, signature of non-equilibrium effect in quantum critical transport. This deviation from the equilibrium power-law is generated by the steady-state non-equilibrium current at large voltage bias which leads to different transport property from that due to equilibrium thermal effect \cite{ChungPRL09}. The conductance $G(V,T)$ via Eq. (\ref{eq:1st ord green}) offers an analytical and complete universal crossover function from 2CK non-equilibrium quantum critical ($V\gg T$) to the equilibrium 2CK ($V\ll T$) limits, which shows $V/T$ scaling (see Fig. \ref{fig:fig1}(d)). Our analytic crossover function in $I$-$V$ curve provides not only a qualitative but also a quantitative basis to compare with experiments. The analytic form in Eq. (\ref{eq:RG-eq}) reduces to a constant conductance for $K=1/2$ in the wide-band ($D\rightarrow\infty$) limit \cite{Shah1}.
 
\textit{\textbf{Conclusions.}}---A few remarks are made before we conclude. First, in Refs. \cite{Mebrahtu12, Mebrahtu13}, the emulated Luttinger wire was realized experimentally in a spin polarized carbon nano-tube quantum dot sbject to an Ohmic dissipation where the resistance $R$ is side-coupled to the dot. The effective Luttinger parameter $K$ is related to the dimensionless dissipation strength $r\equiv Re^2/h$ via $K=1/(1+r)$. When the dot is symmetrically coupled to the leads, the system approaches to a quantum critical point of the 2CK type. The conductance reaches the unitary limit of $G(V,T)\rightarrow2e^2/h$ for $V$, $T\rightarrow0$ in a power law fashion: $G(V,T)\sim(V/T)^{\alpha}$ with $\alpha=2/(1+r)$. Generalizing this set up to the spinful case, the Kondo-Luttinger system equivalent to our model was proposed \cite{FlorensPRB2007} and has been realized experimentally in a dissipative Kondo dot system with $K=1/(1+2r)$ \cite{Finkelstein}. Though their Hamiltonian is somewhat different from Eq. (\ref{eq:KL-hamiltonian}), the same 1CK-2CK QPT occurs at $r=1$ (or $K=1/2$) \cite{FlorensPRB2007}. 

  Secondly, we further included the 2-loop order corrections to Eq. (\ref{eq:RG-eq}) as $\frac{dj_{LR}}{dl}=(1-\frac{1}{2K})j_{LR}-\frac{1}{4}(j_{LR})^3-\frac{1}{8}(\delta j^z)^2j_{LR}$, and $\frac{d\delta j^z}{dl}=-\frac{1}{2K}\delta j^z-\frac{1}{4}(j_{LR})^2\delta j^z-\frac{1}{8}(\delta j^z)^3$ with $j_{LR}=J_{LR}/2\pi\hbar v_F$ and $\delta j^z=\delta J_z/2\pi\hbar v_F$ \cite{suppl}. For $K>1/2$, instead of flowing to a strong coupling 1CK fixed point up to 1-loop order, the $j_{LR}$ term flows to an intermediate coupling 1CK fixed point at $j^{*2}_{LR} = 2(1-\frac{1}{2K})$. For $K<1/2$, there is no new critical point appearing as the linearized RG equations near the 2CK fixed point ($j^*_{LR}=\delta j^{z*}=0$) reduces to the same QCP via Eq. (\ref{eq:RG-eq}). Since $\delta j^z$ term is more irrelevant than $j_{LR}$, a finite $\delta j^z$ will only lead to negligible 2-loop RG corrections to $j_{LR}$ and to the current. The RG flows for $j_{LR}$ (see inset of Fig. 1(c)) show a negligible difference between results up to 1-loop order in the Toulouse limit and 2-loop order away from this limit. Consequently, our results can be extended to parameter regime away from Toulouse limit with a finite $\delta j^z$. Thirdly, channel asymmetric $J_-$ term is a relevant perturbation of our results, making the 2CK fixed point unstable towards the one-lead dominated 1CK fixed point. Nevertheless, the channel symmetry has been achieved experimentally for our model in Ref. \cite{Finkelstein} and for its spinless version in Refs. \cite{Mebrahtu12, Mebrahtu13} via gate-tuning. Finally, in the presence of particle-hole asymmetry, a potential scattering term $U\cos(\phi_{sf}(0))\cos(\phi_{f}(0)/\sqrt{K})$ is generated \cite{GogolinPRB1995, E.Kim}. For $K<1/2$, this new term becomes irrelevant ($[U]=(1+1/K)/2$) and can be neglected; while for $K>1/2$, it is a relevant perturbation ($[U]=(1+K)/2$), and the conducting 1CK state becomes unstable towards the insulating 1CK state ($[J_{LR}]=(1+1/K)/2$) with $G(T)\sim T^{1/K-1}$ as $T\rightarrow0$.

  We have established a novel framework to investigate the non-equilibrium transport near the strong coupling fixed point of a Kondo-Luttinger system close to the well-known one-channel Kondo to two-channel Kondo quantum critical point at the Luttinger parameter $K=1/2$. Via bosonization and re-fermionization of the model near strong coupling two-channel Kondo fixed point, the system is mapped onto an effective anisotropic dissipative Kondo model in the weak coupling regime. Via renormalization group analysis, we identify this quantum critical point. By Keldysh Green's function approach to the effective model, we obtain an analytical form for the non-equilibrium current and conductance near this critical point for $K<1/2$ perturbatively. The interactions in the leads are treated exactly by bosonization, and the current is computed perturbatively in the Kondo coupling. Our results provide an unique example of analytically solvable universal non-equilibrium transport near a quantum critical point in Kondo-Luttinger system. Further experimental investigations in a dissipative Kondo impurity in quantum dot devices is needed to clarify our predictions.
  
  This work is supported by the MOSTGrant No. 104-2112- M-009-004-MY3, the MOE-ATU 392 program, the NCTS of Taiwan, R.O.C. (C.-H. C.).


\begin{thebibliography}{36}
\bibitem {footnote}%
  \BibitemOpen
  \bibfield  {title}  {\bibinfo {title} {In deriving Eq. (7), we have used the identity $\langle c(t)e^{i\phi(t)}c(0)^{\dagger}e^{-i\phi(0)}\rangle=g(t)b(t)$ with $b(t)=\langle e^{i\phi(t)}e^{-i\phi(0)}\rangle_0$, which separates the fermionic and bosonic correlation functions.}}
\end{thebibliography}
\end{document}